\documentclass[12pt]{article}
\usepackage{latexsym}
\usepackage{amsmath}
\usepackage{amssymb}
\catcode`\@=11
\newcount\hour
\newcount\minute
\newtoks\amorpm \hour=\time\divide\hour by 60\minute
=\time{\multiply\hour by 60 \global\advance\minute by-\hour}
\edef\standardtime{{\ifnum\hour<12 \global\amorpm={am}%
        \else\global\amorpm={pm}\advance\hour by-12 \fi
        \ifnum\hour=0 \hour=12 \fi
        \number\hour:\ifnum\minute<10
        0\fi\number\minute\the\amorpm}}
\edef\militarytime{\number\hour:\ifnum\minute<10
0\fi\number\minute}
\def\draftlabel#1{{\@bsphack\if@filesw {\let\thepage\relax
   \xdef\@gtempa{\write\@auxout{\string
      \newlabel{#1}{{\@currentlabel}{\thepage}}}}}\@gtempa
   \if@nobreak \ifvmode\nobreak\fi\fi\fi\@esphack}
        \gdef\@eqnlabel{#1}}
\def\@eqnlabel{}
\def\@vacuum{}
\def\marginnote#1{}
\def\draftmarginnote#1{\marginpar{\raggedright\scriptsize\tt#1}}
\def\draft{
        \pagestyle{plain}
        \overfullrule=2pt
        \oddsidemargin -.1truein
        \def\@oddhead{\sl \phantom{\today\quad\militarytime} \hfil
        \smash{\Large\sl DRAFT} \hfil \today\quad\militarytime}
        \let\@evenhead\@oddhead
        \let\label=\draftlabel
        \let\marginnote=\draftmarginnote
        \def\ps@empty{\let\@mkboth\@gobbletwo
        \def\@oddfoot{\hfil \smash{\Large\sl DRAFT} \hfil}
        \let\@evenfoot\@oddhead}
        \def\@eqnnum{(\theequation)\rlap{\kern\marginparsep\tt\@eqnlabel}%
        \global\let\@eqnlabel\@vacuum}  }

\renewcommand{\thefootnote}{\fnsymbol{footnote}}

\def\appendix#1{\addtocounter{section}{1}\setcounter{equation}{0}
\renewcommand{\thesection}{\Alph{section}}
\section*{Appendix \thesection\protect\indent \parbox[t]{11.15cm}{#1}}
\addcontentsline{toc}{section}{Appendix \thesection\ \ \ #1}}

\def \bi{\bibitem}
\def \la {\label}

\def\be{\begin{equation}}
\def\ee{\end{equation}}

\catcode`\@=11

\def\bea{\begin{eqnarray}}
\def\eea{\end{eqnarray}}
\def\beann{\begin{eqnarray*}}
\def\eeann{\end{eqnarray*}}
\def\beq{\begin{equation}}
\def\eeq{\end{equation}}
\def\ba{\begin{array}}
\def\ea{\end{array}}
\def\ben{\begin{enumerate}}
\def\een{\end{enumerate}}

 \def \la {\label}
 \def\be{\begin{equation}}
\def\ee{\end{equation}}

\def \la {\label}

\font\mybb=msbm10 at 11pt

\def\bb#1{\hbox{\mybb#1}}

\def\bR {\bb{R}}

\def\bC {\bb{C}}

\def \ee {\epsilon}

\def \bi{\bibitem}
\def\a{\alpha }

\def\be{\begin{equation}}
\def\ee{\end{equation}}

\def \bi {\bibitem}
\def \la{\label}

\begin{document}
\date{November 2002}
\begin{titlepage}
\begin{center}
\hfill hep-th/0606049 \\
\hfill KUL-TF-06/20 \\

\vspace{3.0cm} {\Large \bf N=31 is not IIB}
\\[.2cm]

\vspace{1.5cm}
 {\large  U. Gran$^1$, J. Gutowski$^2$,  G. Papadopoulos$^3$ and D. Roest$^3$}

\vspace{0.5cm}

${}^1$ Institute for Theoretical Physics, K.U. Leuven\\
Celestijnenlaan 200D\\
B-3001 Leuven, Belgium\\

\vspace{0.5cm}
${}^2$ DAMTP, Centre for Mathematical Sciences\\
University of Cambridge\\
Wilberforce Road, Cambridge, CB3 0WA, UK

\vspace{0.5cm}
${}^3$ Department of Mathematics\\
King's College London\\
Strand\\
London WC2R 2LS, UK\\
\end{center}

\vskip 1.5 cm
\begin{abstract}

We  adapt the spinorial geometry method  to investigate supergravity backgrounds
with near maximal number of supersymmetries. We then apply the formalism to
show that the IIB supergravity backgrounds with 31 supersymmetries preserve an additional supersymmetry and
so they are maximally supersymmetric. This rules out the existence of IIB supergravity preons.

\end{abstract}
\end{titlepage}
\newpage
\setcounter{page}{1}
\renewcommand{\thefootnote}{\arabic{footnote}}
\setcounter{footnote}{0}

\setcounter{section}{0}
\setcounter{subsection}{0}

It has been known for some time that a priori in type II and
eleven-dimensional supergravities there may exist  backgrounds with
any number of supersymmetries. This is because the holonomy of the
supercovariant connection of these theories is a subgroup of
$SL(32,\bR)$ and so any $N<32$ spinors have a non-trivial stability
subgroup in the holonomy group. For a more detailed  explanation see
\cite{hull, duffl, gpta} for the M-theory and \cite{gptb} for IIB.
Furthermore, it was argued in \cite{ugjggpdr} that the Killing
spinor bundle ${\cal K}$ can be any subbundle of the Spin bundle and
the spacetime geometry depends on the trivialization of ${\cal K}$.
This is unlike what happens in the case of Riemannian and Lorentzian
geometries \cite{berger, figueroab} and heterotic and type I
supergravities\footnote{This is provided the parallel spinors are
Killing.} \cite{gpgran}, where there are restrictions both on the
number of Killing spinors and the Killing spinor bundle.

In this paper, we shall show that IIB backgrounds with 31
supersymmetries are maximally supersymmetric. Backgrounds with 31
supersymmetries have been considered before in the context of
M-theory  \cite{bandos} and have been termed as preons. To our
knowledge this is the first example which demonstrates that there
are restrictions on the number of supersymmetries of type II
backgrounds. To do this, we shall adapt the spinorial method
\cite{jguggp} of solving Killing spinor equations to backgrounds
that admit near maximal number of supersymmetries. We shall mostly
focus on IIB  and eleven-dimensional supergravity but most of
the analysis extends to all supergravity theories.

To adapt the spinorial method to backgrounds with near maximal number of supersymmetries, we
 introduce   a ``normal'' ${\cal K}^\perp$ to the Killing spinor bundle ${\cal K}$ of a supersymmetric background.
 The spinors of IIB supergravity
are complex positive chirality Weyl spinors, so the Spin bundle is
${\cal S}^c_+={\cal S}_+\otimes\bC$, where ${\cal S}_+$ is the rank
sixteen  bundle of positive chirality Majorana-Weyl spinors. ${\cal
S}^c_+$  may also be thought of as an associated bundle of a
principal bundle with fibre $SL(32, \bR)$, the holonomy group of the
supercovariant connection, acting with the fundamental
representation on $\bR^{32}$. If a background admits $N$ Killing
spinors which span the fibre of the Killing spinor bundle ${\cal
K}$, then one has the sequence \bea 0\rightarrow {\cal K}\rightarrow
{\cal S}_+^c\rightarrow {\cal S}_+^c/{\cal K}\rightarrow 0~.
\la{exseq} \eea The inclusion   $i:\,{\cal K}\rightarrow {\cal
S}_+^c$ can be locally described as \bea \epsilon_r=\sum_{i=1}^{32}
f^i{}_r \eta_i~,~~~r=1,\dots, N \,,\eea where $\eta_p$,
$p=1,\dots,16$, is a basis in the space of positive chirality
Majorana-Weyl spinors, $\eta_{16+p}= i\eta_p$ and the coefficients
$f$ are real spacetime functions. For our notation and spinor
conventions see \cite{ugjggpdr}. Any $N$  Killing spinors related by
a local $Spin(9,1)$ transformation give rise to the same spacetime
geometry. This is because the Killing spinor equations and the field
equations of IIB supergravity are Lorentz invariant. Therefore any
bundles of Killing spinors and any choice of sections related by a
$Spin(9,1)$ gauge transformation\footnote{IIB supergravity has a
$Spin(9,1)\times U(1)$ gauge symmetry but the restriction to
$Spin(9,1)$ will suffice.}
  should be identified.

To construct ${\cal K}^\perp$, first consider the dual ${}^\star{\cal S}_+^c$ of ${\cal S}_+^c$ and
introduce a basis $\eta^i$, $\eta^i(\eta_j)=\delta^i{}_j$, i.e. $\eta^{16+p}=-i\eta^p$.
Next consider the sections $\a$ of ${}^\star{\cal S}_+^c$ that annihilate the Killing spinors $\epsilon_r$, i.e
$\a(\epsilon)=0$, or equivalently
\bea
f^i{}_r u_i=0~,~~~~\a=u_i \eta^i~,
\eea
where $u_i$ are real spacetime functions. Since  the matrix $f=(f^i{}_r)$ has rank $N$, there are $32-N$ solutions
to this equation. These solutions span the sections of the co-kernel, ${\rm coker}\, i\subset {}^\star{\cal S}^c_+$
 of the inclusion map
$i: {\cal K}\rightarrow {\cal S}_+^c$. It is well-known that $Spin(9,1)$ has an invariant
inner product $B: {\cal S}_+\otimes {\cal S}_-\rightarrow \bR$
\bea
B(\epsilon, \zeta)=-B(\zeta, \epsilon)=<B(\epsilon^*), \zeta>~,
\eea
which extends to $B: {\cal S}^c_+ \otimes {\cal S}^c_-\rightarrow
\bC$ as a bi-linear in both entries. Next consider
\bea
{\cal B}(\epsilon, \zeta)={\rm Re}\, B(\epsilon, \zeta)~,
\eea
which defines a non-degenerate pairing ${\cal B}: {\cal S}^c_+
\otimes {\cal S}^c_-\rightarrow \bR$. This in turn induces a
isomorphism $j: {}^\star{\cal S}^c_+\rightarrow {\cal S}^c_-$ as
${\cal B}(j(\a), \epsilon)=\a(\epsilon)$. We identify the image of
$j$,  $j({\rm coker}\, i)\subset {\cal S}^c_- $, as the ``normal''
bundle ${\cal K}^\perp$ of ${\cal K}$, i.e. $j({\rm coker}\,
i)={\cal K}^\perp$. Clearly if $\a\in {\rm coker}\, i$ and
$\epsilon\in {\cal K}$, then $\a(\epsilon)=0$, and so one gets the
``orthogonality'' condition,
\bea
{\cal B}(j(\a), \epsilon)=0~.
\la{normcol}
\eea
Observe that ${\cal S}_+^c/{\cal K}={}^\star{\cal K}^\perp$.
To write this orthogonality condition in components, introduce  a basis in ${\cal S}^c_-$, say $\theta_{i'}=-\Gamma_0\eta_i$. Then
write $j(\a)=\nu=n^{i'} \theta_{i'}$ and the condition (\ref{normcol}) can be written as
\bea
n^{i'} {\cal B}_{i' j} f^j{}_r=0~,
\la{normcolcom}
\eea
where ${\cal B}_{i'j}={\cal B}(\theta_{i'}, \eta_j)$.

The condition (\ref{normcol}), or equivalently (\ref{normcolcom}),
leads to a correspondence
 between the $N$ Killing spinors and the $32-N$ normal directions, i.e.
\bea
N\longleftrightarrow 32-N~.
\eea
 This is because instead of specifying
the Killing spinors, one can  determine the normal spinors. Substituting the normal
spinors into these equations, one can then solve for
 the Killing spinors.
In addition, the construction of ${\cal K}^\perp$ and  (\ref{normcol}) or (\ref{normcolcom}) are
$Spin(9,1)$ covariant. Because of this, the $Spin(9,1)$ gauge symmetry can be used
to bring the normal spinors instead of the Killing spinors into a canonical form. In turn, this leads to a simplification
in the expression for the Killing spinors which can be used to solve
the Killing spinor equations for backgrounds with near maximal number of supersymmetries. We shall demonstrate this
for IIB backgrounds with 31 supersymmetries.
Furthermore, one may
 consider cases such that the sections of ${\cal K}^\perp$ are invariant under some non-trivial stability
subgroup of $Spin(9,1)$. It is clear these cases are related to (e.g.~maximal and half-maximal) $G$-backgrounds
 \cite{ugjggpdr, gmaxsusy}, where the invariance  condition was imposed on the Killing spinors.
 The  spinorial geometry techniques  that we use  to investigate backgrounds with $N$
 supersymmetries can be adapted to examine backgrounds with $32-N$ supersymmetries and vice-versa.

One can easily extend the construction described above to M-theory. In particular, one again has
\bea
0\rightarrow {\cal K}\rightarrow {\cal S}\rightarrow {\cal S}/{\cal
K}\rightarrow 0 \,,
\eea
where ${\cal S}$ is the spin bundle associated with the Majorana representation of $Spin(10,1)$. The inclusion map
$i:\, {\cal K}\rightarrow {\cal S}$ can be written locally as $\epsilon_r= \sum_{i=1}^{32} f^i{}_r \eta_i$, where $f^i{}_r$ are real
spacetime functions and $(\eta_i, i=1, \dots, 32)$ is a basis of Majorana spinors. As in the IIB case, we consider the
the co-kernel of the inclusion map $i:\, {\cal K}\rightarrow {\cal S}$, ${\rm coker}\, i\subset {}^\star {\cal S}$.
It is well known that ${\cal S}$ admits a $Spin(10,1)$ invariant
inner product $B$ which gives rise to an isomorphism $j:\, {}^\star {\cal S}\rightarrow {\cal S}$. As in the IIB
case, we define the normal bundle of the Killing spinor bundle as ${\cal K}^\perp= j({\rm coker}\, i)$.
In this case, ${\cal K}^\perp$ is a subbundle of ${\cal S}$ and ${\cal S}/{\cal K}={\cal K}^\perp$.
Taking a section $\nu= n^i\eta_i$ of ${\cal K}^\perp$,
the orthogonality condition analogous to (\ref{normcol}) and (\ref{normcolcom}) is
\bea
n^i B_{ij} f^j{}_r=0~,
\la{elcon}
\eea
where $B_{ij}=B(\eta_i, \eta_j)$. The condition (\ref{elcon}) is $Spin(10,1)$ covariant.

As an example consider IIB backgrounds that admit 31 supersymmetries. According to the correspondence $N\leftrightarrow 32-N$,
these are related to backgrounds
with one supersymmetry investigated in \cite{ugjggpa, ugjggpdr}. To carry out the computation,
we need to find the canonical form of spinors in ${\cal S}^c_-$ up to $Spin(9,1)$ transformations.
It is easy to deduce using an argument similar to \cite{ugjggpa} that there are three kinds of orbits of $Spin(9,1)$
in the negative chirality Weyl spinors with stability subgroups $Spin(7)\ltimes\bR^8$, $SU(4)\ltimes \bR^8$ and
$G_2$. A canonical form of these spinors is
\bea
&&\nu_1=(n+im) (e_5+e_{12345})~,~~~\nu_2= (n-\ell+im) e_5+ (n+\ell+im) e_{12345}~,
\cr
&&\nu_3=n(e_5+e_{12345})+i m (e_1+e_{234})~,
\eea
respectively.
Using the $Spin(9,1)$ gauge symmetry, we choose ${\cal K}^\perp$ to lie along the directions of one of the
above spinors. Consider first the $\nu_1$ case. Write the
Killing spinors as
\bea
\epsilon_r= f^1{}_r (1+e_{1234})+ f^{17}{}_r i (1+e_{1234})+ f^k{}_r
\eta_k \,,
\eea
where $\eta_k$ are remaining basis elements complementary to $1+e_{1234}$ and $i(1+e_{1234})$.
In what follows, we use the basis constructed from the five types of spinors in \cite{ugjggpdr}.
Substituting $\epsilon_r$ into  (\ref{normcol}), we
get
\bea
f^1{}_r n- f^{17}{}_r m=0~.
\eea
Without loss of generality, we take  $n\not=0$. Using this,  we solve for $f^1{}_r$ and substitute back
into the Killing spinors to find
\bea
\epsilon_r={f^{17}{}_r\over n} (m+in) (1+e_{1234})+ f^k{}_r \eta_k~.
\eea
Similarly for the normal spinors $\nu_2$ and $\nu_3$, we find that
\bea
&&\epsilon_r={f^{17}{}_r\over n}[
(m+in)(1+e_{1234})]+{f^{18}{}_r\over n} [\ell (1+e_{1234})-n
(1-e_{1234})]+ f^{k}{}_r \eta_{k} \,,
\cr
&&\epsilon_r={f^{19}{}_r\over n} [m (1+e_{1234})+i n(e_{15}+e_{2345})]+ f^k{}_r \eta_k~,
\eea
correspondingly, where $\eta_k$ are the remaining basis elements in each case.
Substituting these spinors into the algebraic Killing spinor equation and using that the rank of
the matrix $(f^i{}_r)$ is 31,  for the $Spin(7)\ltimes \bR$ case one finds that
\bea
&&P_M\Gamma^M C*[(m+in) (1+e_{1234})]+{1\over24} G_{M_1M_2M_3}\Gamma^{M_1M_2M_3} (m+in) (1+e_{1234})=0~,
\cr
&&P_M\Gamma^M\eta_p=0~,~~~G_{M_1M_2M_3}\Gamma^{M_1M_2M_3}\eta_p=0~,~~~p=2,\dots, 16~,
\la{cona}
\eea
and similarly
\bea
&&P_M\Gamma^M C*[(m+in) (1+e_{1234})]+{1\over24} G_{M_1M_2M_3}\Gamma^{M_1M_2M_3} (m+in) (1+e_{1234})=0~,
\cr
&&P_M\Gamma^M C*[\ell  (1+e_{1234})-n (1-e_{1234})]
\cr
&&~~~~~~~~~~+{1\over24} G_{M_1M_2M_3}\Gamma^{M_1M_2M_3} [\ell (1+e_{1234})-n (1-e_{1234})]=0~,
\cr
&&P_M\Gamma^M C*[i (1-e_{1234})]+{1\over24} G_{M_1M_2M_3}\Gamma^{M_1M_2M_3} [i (1-e_{1234})]=0~,
\cr
&&P_M\Gamma^M\eta_{p}=0~,~~~G_{M_1M_2M_3}\Gamma^{M_1M_2M_3}\eta_{p}=0~,~~~p=3,\dots,16~,
\la{conb}
\eea
and
\bea
&&P_M\Gamma^M C*[m (1+e_{1234})+in(e_{15}+e_{2345})]
\cr
&&~~~~~~~~~~~~~~+{1\over24} G_{M_1M_2M_3}\Gamma^{M_1M_2M_3} [m (1+e_{1234})+in(e_{15}+e_{2345})]=0~,
\cr
&&P_M\Gamma^M C*(i (1+e_{1234})+{1\over24} G_{M_1M_2M_3}\Gamma^{M_1M_2M_3} (i (1+e_{1234})=0~,
\cr
&&P_M\Gamma^M C* (e_{15}+e_{2345})+{1\over24} G_{M_1M_2M_3}\Gamma^{M_1M_2M_3} (e_{15}+e_{2345})=0~,
\cr
&&P_M\Gamma^M\eta_p=0~,~~~G_{M_1M_2M_3}\Gamma^{M_1M_2M_3}\eta_p=0~,~~~p=2,4,\dots,16~,
\la{conc}
\eea
for the other two cases. The factorization of $P$ and $G$ flux terms on $\eta_p$  occurs because some of the
remaining basis elements $\eta_k$
 come in complex conjugate pairs $(\eta_p, i\eta_p)$, where $\eta_p$ are Majorana-Weyl spinors. Since
 the  $P$ flux term in the Killing spinor equations contains the charge conjugation matrix, $C*\eta_p=\eta_p$ and
 $C*(i \eta_p)=-i\eta_p$, there is a relative sign between the $P$ and $G$ flux terms when the algebraic Killing spinor
 equation is evaluated on
 $\eta_p$ and $i\eta_p$. It now remains to solve these equations.

First, focus on the equation $P_M \Gamma^M\eta_p=0$. Observe
that in all cases, the remaining spinors $\eta_p$ contain spinors which are annihilated by either
$\Gamma^-$ or $\Gamma^+$. In the former case,  the condition $P_M\Gamma^M\eta_p=0$ implies that only the $P_-$
component is non-vanishing
  while in the latter case implies that only the component $P_+$ is non-vanishing. Since spinors of both types occur,  $P=0$.

Next consider the conditions on the $G$ flux. It turns out that  (\ref{cona}), (\ref{conb}) or (\ref{conc}) imply
that  $G_{M_1M_2M_3}\Gamma^{M_1M_2M_3}\epsilon=0$ for all spinors
$\epsilon$ and so  $G=0$. To see this consider the $Spin(7)\ltimes\bR^8$ case. Setting $P=0$ in the first condition
in (\ref{cona}), we deduce that
$G_{M_1M_2M_3}\Gamma^{M_1M_2M_3}(1+e_{1234})=0$. Since the algebraic Killing spinor equations with $P=0$
are linear over the complex numbers, we also have that
$G_{M_1M_2M_3}\Gamma^{M_1M_2M_3}i(1+e_{1234})=0$. This together with the remaining conditions in (\ref{cona}) imply
that $G_{M_1M_2M_3}\Gamma^{M_1M_2M_3}\eta_i=0$ for all the basis elements $\eta_i$. A similar argument applies to the rest of the
cases. Thus we have found that the algebraic Killing spinor equations imply that $P=G=0$. We have also verified this by
an explicit computation.

Finally, if the $P$ and $G$ fluxes vanish, then the gravitino Killing spinor equation
of IIB supergravity becomes linear over the complex numbers. This means that backgrounds with vanishing
$P$ and $G$ fluxes always preserve an even number of supersymmetries. Thus backgrounds with 31
supersymmetries preserve an additional supersymmetry and so they
are maximally supersymmetric. In particular, they  are
 locally isometric \cite{jfgpa} to Minkowski spacetime, $AdS_5\times S^5$ \cite{schwarz} and the maximally supersymmetric plane wave
\cite{bfhp}. As a corollary, we have shown that  IIB supergravity preons do not exist.

Our proof has relied  on the algebraic Killing spinor equation of IIB supergravity and so does not straightforwardly generalize
to eleven-dimensional supergravity. Nevertheless, as we have seen the normal Killing spinor bundle construction generalizes to M-theory.
In addition,
one can show that the 31 Killing spinors of M-theory preon backgrounds take a simple form and it may
be possible to solve the Killing spinor equations. We hope to report on the existence
of M-theory preons in the  future.

\vskip 0.5cm

\noindent{\bf Acknowledgements} \vskip 0.1cm The research of D.R.~is
funded by the PPARC grant PPA/G/O/2002/00475 and U.G.~has a
postdoctoral fellowship funded by the Research Foundation
K.U.~Leuven.

\end{document}